# Enhancing Power Prediction of Photovoltaic Systems: Leveraging Dynamic Physical Model for Irradiance-to-Power Conversion

Baojie Li, Xin Chen, Anubhav Jain
Energy Technologies Area, Lawrence Berkeley National Laboratory, Berkeley, CA, USA

## Abstract

Power prediction is crucial to the efficiency and reliability of Photovoltaic (PV) systems. For the model-chain-based (also named indirect or physical) power prediction, the conversion of ground environmental data (plane-of-array irradiance and module temperature) to the output power is a fundamental step, commonly accomplished through physical modeling. The core of the physical model lies in the parameters. However, traditional parameter estimation either relies on datasheet information that cannot reflect the system's current health status or necessitates additional I-V characterization of the entire array, which is not commonly available. To address this, our paper introduces PVPro, a dynamic physical modeling method for irradiance-to-power conversion. It extracts model parameters from the recent production data without requiring I-V curve measurements. This dynamic model, periodically-updated (as short as daily), can closely capture the actual health status, enabling precise power estimation. To evaluate the performance, PVPro is compared with the smart persistence, nominal physical, and various machine learning models for day-ahead power prediction. The results indicate that PVPro achieves an outstanding power estimation performance with the average $nMAE$ =1.4% across four field PV systems, reducing the error by 17.6% compared to the best of other techniques. Furthermore, PVPro demonstrates robustness across different seasons and weather conditions. More importantly, PVPro can also perform well with a limited amount of historical production data (3 days), rendering it applicable for new PV systems. The tool is available as a Python package at: https://github.com/DuraMAT/pvpro.

| Nomenclature | | | |
|---|---|---|---|
| $CV$ | Coefficient of variation | ANN | Artificial neural network |
| $GHI$ | Global horizontal irradiance (W/m$^2$) | c-Si | Crystalline silicon |
| $G_{POA}$ | Global plane of Array irradiance (W/m$^2$) | DC | Direct current |
| $I$ | Current (A) | DST | Daylight saving time |
| $I_o$ | Saturation current (A) | PV | Photovoltaic |
| $I_{ph}$ | Photocurrent (A) | I-V curve | Current-voltage characteristic |
| $I_{DC}$ | DC current (A) | KR | Kernel Ridge |
| $I_{sc}$ | Short-circuit current (A) | LR | Linear regression |
| $n$ | Diode factor | ML | Machine learning |
| $nMAE$ | Normalized mean absolute error | MLP | Multilayer Perceptron |
| $nRMSE$ | Normalized root mean square error | MPP | Maximum power point |
| $nBE$ | Normalized bias error | NWP | Numerical weather prediction |
| | | PV | Photovoltaic |
| $P$ | Power (W) | RF | Random forest |
| $R_s$ | Series resistance (Ω) | SDM | Single diode model |
| $R_{sh}$ | Shunt resistance (Ω) | STC | Standard test condition |
| $V$ | Voltage (V) | SVR | Support vector regression |
| $V_{DC}$ | DC voltage (V) | | |
| $V_{oc}$ | Open-circuit voltage (V) | | |

## 1. Introduction

Photovoltaic (PV) power prediction plays a significant role in optimizing the efficiency and reliability of solar energy systems [1,2]. Anticipating the power output of photovoltaic (PV) systems enables effective energy management, aiding in grid integration and balancing electricity supply and demand. Even a one percent change in power prediction accuracy holds considerable electrical market value [3].

The flowchart of typical PV power prediction is illustrated in Fig. 1 (a). Weather forecasting is the first and essential step [2], where the weather data can be forecasted from numerical weather prediction (NWP) [4], sky cameras [5], or satellite images [6]. The next step is the Irradiance-to-Power conversion. Note that the 'irradiance' here generally refers to the Global Horizontal Irradiance ($GHI$) obtained by NWP or sky/satellite images [7]. The common Irradiance ($GHI$)-to-Power conversion methods can be broadly categorized into direct (or statistical) and indirect (or physical model chain) approaches [8,9], as depicted in Fig. 1. Direct prediction methods leverage advanced techniques like machine learning or classical statistical algorithms (like regressive methods) [1,9]. These approaches analyze historical data to identify patterns and correlations, enabling the prediction of power output based on past system behavior. The indirect ones involve a multi-step model chain [8,10] to compute the power. An extensive review of the model chain can be found in [8]. In Fig. 1 (a), we simplify this model chain into two steps. The first is the post-processing of the forecasted weather data, which includes tasks such as the separation of the beam and diffuse components from $GHI$ [11], transposition of the global beam and diffuse irradiance to the plane-of-array irradiance ($G_{poa}$) [12], and the estimation of the module temperature from the forecasted ambient temperature and wind speed [13]. Then, the computed ground weather data ($G_{poa}$ and module temperature) are converted to the output power, a process also referred to as Irradiance-to-Power conversion (or 'PV performance models' [8]). In this context, the term 'irradiance' specifically pertains to the irradiance reaching the inclined surface of the PV module, *i.e.*, $G_{poa}$ [8].

There are various Irradiance ($G_{poa}$) -to-Power conversion methods, as presented in Fig. 1 (b), including modeling via equivalent physical models [14], machine learning models [15], or persistence models [16]. Machine learning models offer adaptability to complex non-linear patterns, learning from data-driven insights for accurate predictions, and accommodating dynamic changes over time [17].

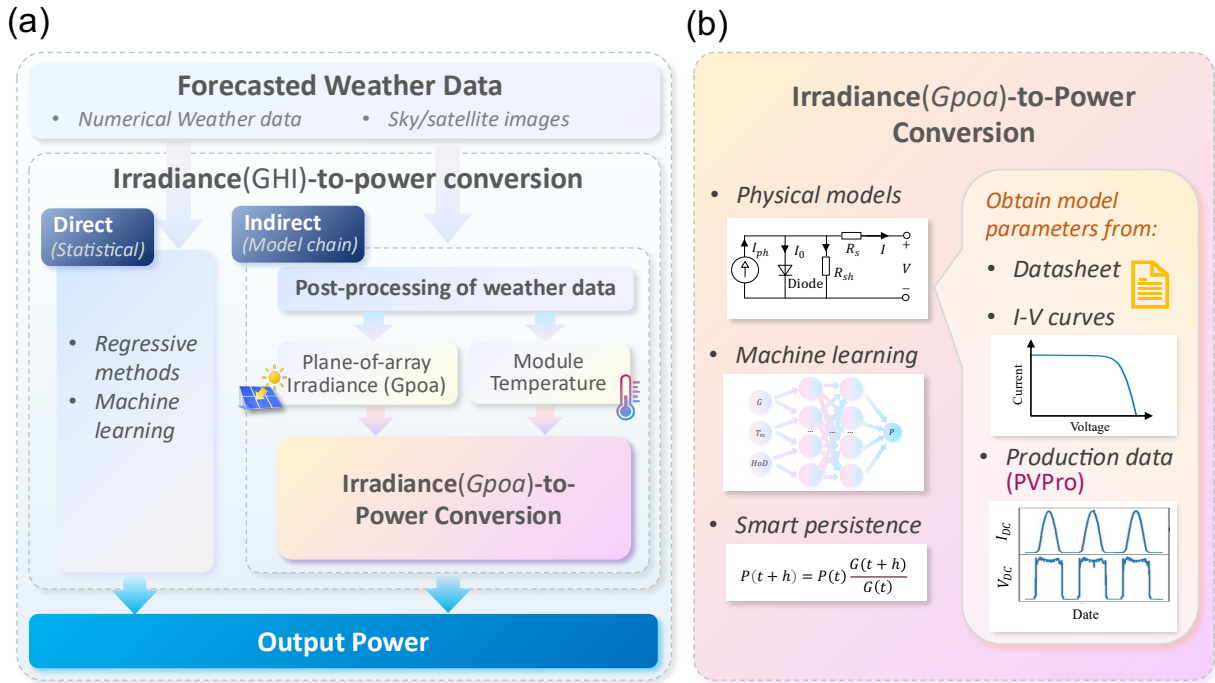

Fig. 1 (a) Flowchart of the common PV power forecast techniques. Two main categories of irradiance-to-power conversion techniques, direct (statistical) and indirect (model chain), map the forecasted weather data to the output power. The direct ones typically employ machine learning and classical statistical models. The indirect ones follow a physical model chain to predict the power. (b) Common Irradiance ($G_{poa}$) -to-power conversion methods (physical modeling, machine learning, and smart persistence model). The physical model parameters can be obtained from datasheets, I-V curves, or production data (DC voltage and current). PVPro leverages the production data to rebuild the physical model for the Irradiance ($G_{poa}$) -to-power conversion.

Physical equivalent circuit modeling is the most common method for the irradiance ($G_{poa}$) -to-power conversion within the model chain [1]. The typical physical models are the single-, double-, or three-diode models, distinguished by the number of parameters used to characterize the model [18]. The number of parameters in these models can be three [19], four [20], five [21], six [22], or seven [23]. A review of these different physical models can be found in [14] [24]. After the literature study, we identified that the major challenge of physical modeling for power prediction lies in the accurate estimation of physical model parameters [25]. The parameter estimation using nominal manufacture datasheet information poses a significant limitation [8], as they cannot accurately reflect the current health status of PV systems, particularly those affected by unknown faults or years of degradation [26]. Using field-measured current-voltage characteristics (I-V curves) enables the acquisition of high-precision physical parameters [27]. However, the characterization of field I-V curves necessitates additional measurement equipment and will interrupt the operation of the system. Thus, the I-V curves of the entire PV array are not readily available, especially for large-scale PV systems [18]. While it is feasible to characterize a single reference PV module installed near the array, the health status of this reference module may not fully represent that of the entire PV array [28]. These findings emphasize the complexity and limitations in the parameter estimation of physical models for PV power prediction.

Seen in this light, this paper proposes a dynamic physical modeling method, PVPro, to perform irradiance-to-power conversion for PV power prediction. PVPro is a tool we proposed to extract the single-diode model parameters from the recent production data of the PV system without the need for I-V curves [29,30]. Along with the operation of the PV system, the physical model will also be dynamically reconstructed. In this way, the actual health status or the occurrence of faults will be closely captured by this dynamic model. Consequently, a precise output power of the PV system can be modeled. The contribution of this paper is then reflected in the following points: 1) A dynamic-physical-model-based irradiance-to-power conversion method for precise power prediction is proposed; 2) This method is suitable for field PV systems even suffering degradation or faults; 3) The power conversion performance is robust to seasonal and weather impacts; 4) This method is also applicable on newly-installed systems with a limited amount of production data.

The remainder of the paper is organized as: Section 2 outlines the comprehensive methodology, encompassing details on the utilized data, error metrics, and techniques. These techniques include the proposed dynamic physical model, alongside common Irradiance ($G_{poa}$)-to-power conversion methods for comparative analysis, such as the physical model based on nominal parameters, various machine learning models, and the smart persistence model. Section 3 presents the day-ahead Irradiance ($G_{poa}$)-to-power conversion performance across an entire year on four distinct field PV systems. The effects of seasons and weather conditions are specifically analyzed. The over/under-estimation of power and the interpretability of models are also discussed. Section 4 summarizes the pros and cons of the evaluated Irradiance ($G_{poa}$)-to-power conversion techniques. Finally, Section 5 concludes the paper.

## 2. Methodology

This section provides a comprehensive picture of the methodology of the Irradiance($G_{poa}$)-to-power conversion in the model chain, including the data preparation for analysis in Section 2.1, the error metrics to quantify the performance in Section 2.2, and the Irradiance($G_{poa}$)-to-power conversion techniques to evaluate in Section 2.3. The forecasting horizon is set as the day ahead, *i.e.*, 24 hours ahead. This time horizon is crucial for addressing market bidding considerations and ensuring the stability of PV systems [24].

### 2.1. Data

#### 2.1.1. Dataset metadata

We select four PV systems with different capacity scales and diverse climate zones in the U.S. from the PVDAQ data lake [31] as mapped in Fig. 2. The climate zone is determined by the PV Climate Zone (PVCZ) method [32], which distinguish locations based on climate stressors more relevant to PV degradation.

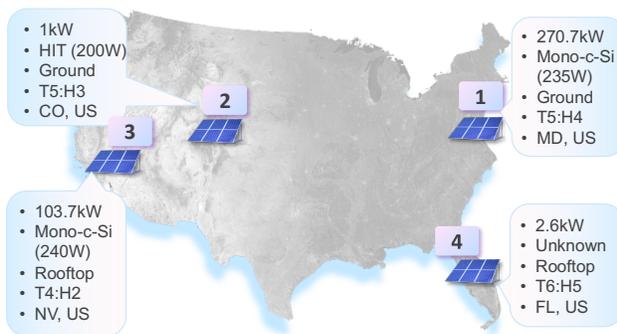

Fig. 2 Location of the four PV systems in the U.S. for evaluation. The metadata of the PV system is listed, including the capacity, module type, mounting type, climate zone, and location.

Note that, to illustrate the detailed process of irradiance-to-power conversion and the different case studies, System 1 is adopted as the primary example. For Systems 2, 3, and 4, the results will be presented in Section 3.2.

### 2.1.2. Pre-processing of data

Preprocessing is critical for ensuring the quality and consistency of PV data for building reliable irradiance-to-power conversion models. Here, two major steps are performed: Daylight saving time (DST) correction and outlier removal. The DST shifts are adjusted using the Solar-data-tools [33]. The outliers are identified by performing a linear regression of the DC current as a function of $G_{poa}$ and of the DC voltage by module temperature as detailed in [29].

As illustrated in Fig. 1, within the model chain of photovoltaic (PV) power prediction, errors are unavoidably introduced during both the weather forecasting and post-processing steps. The primary contributor to power prediction inaccuracies is reported to be the weather forecasting phase [9]. To mitigate errors stemming from these preceding stages, we opt for field-measured environmental data, specifically plane-of-array irradiance ($G_{poa}$) and module temperature, instead of relying on forecasted values. This choice allows for a precise evaluation of the various Irradiance ($G_{poa}$)-to-power conversion techniques.

## 2.2. Error metrics

A single metric is generally insufficient to characterize the power conversion error [9]. In this study, we adopt three common metrics: normalized mean absolute error ($nMAE$), normalized root mean square error ($nRMSE$), and normalized bias error ($nBE$) as described from (1)-(3). All metrics are normalized by the nominal capacity of the PV system.

$$nMAE = \frac{\sum_{i=1}^{N}|P_{pred,i} - P_{meas,i}|}{P_{nominal}} \quad (1)$$

$$nRMSE = \frac{\sqrt{\frac{1}{N}\sum_{i=1}^{N}(P_{pred,i} - P_{meas,i})^2}}{P_{nominal}} \quad (2)$$

$$nBE = \frac{P_{pred,i} - P_{meas,i}}{P_{nominal}} \quad (3)$$

where, $P_{meas}$ and $P_{pred}$ are the measured and predicted power, respectively. $P_{nominal}$ refers the nominal power of the PV system. $nMAE$ reflects the total imbalance between the estimated and the actual power. $nRMSE$ imposes more significant penalties for larger errors. $nBE$ indicates the over- or under-estimation of power.

## 2.3. Irradiance($G_{poa}$)-to-power conversion techniques

This section presents the proposed dynamic physical model-based method. Furthermore, the common Irradiance($G_{poa}$)-to-power conversion techniques for comparison will also be briefly introduced, including traditional physical, smart persistence, and machine learning models.

### 2.3.1. Physical models

Physical models estimate the power by modeling the PV system using equivalent electrical circuit models. Here, we employ the widely-used DeSoto singe-diode model (SDM) [21], which includes five primary parameters, *i.e.*, the photocurrent ($I_{ph}$), saturation current ($I_o$), series resistance ($R_s$), shunt resistance ($R_{sh}$), and the diode factor ($n$).

- **Dynamic physical model (PV-Pro)**

The modeling performance is fundamentally determined by the five parameters. Accordingly, we present a dynamic physical model method, PV-Pro, which determines the model parameters without the need for additional measurements (like I-V curves). The workflow of PV-Pro for irradiance-to-power conversion is illustrated in Fig. 3. PV-Pro only relies on the historical production data (DC voltage ($V_{DC}$) and current ($I_{DC}$)) and weather data (irradiance and module temperature). After an initial guess of the five model parameters based on the module datasheet, PV-Pro models the PV system and gets the simulated $V_{DC}$ and $I_{DC}$. The L2 loss [34] is then calculated between the measured and simulated $V_{DC}$ and $I_{DC}$. Using this loss, L-BFGS-B solver [35] updates the model parameters iteratively until the loss is minimized or the maximum iterations are reached. Further details on this fitting process are elaborated in [29]. This whole process is repeated periodically. The update frequency depends on the user's settings and can be as daily, weekly, monthly, etc. The obtained dynamic physical model parameters can closely catch the PV system's actual health condition and then enhance the accuracy of the irradiance-to-power conversion.

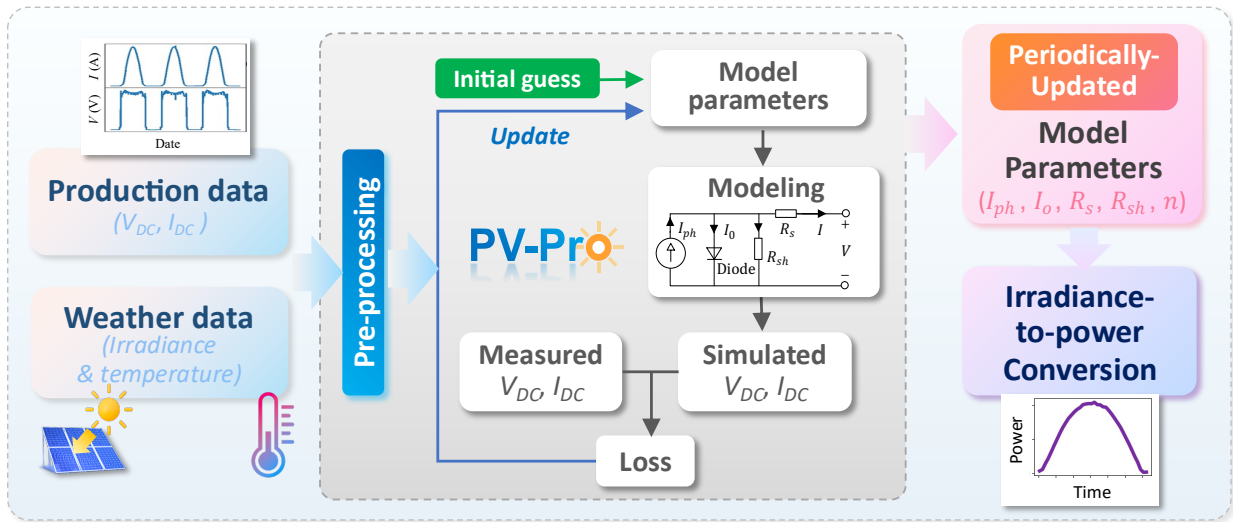

Fig. 3 Flowchart illustrating the estimation of model parameters using PV-Pro for irradiance-to-power conversion. PV-Pro fits the model parameters by minimizing the loss between measured and simulated $V_{DC}$ and $I_{DC}$. Using recent production and weather data, PV-Pro can obtain the periodically updated model parameters that reflect the current health status of the PV system. These dynamic parameters could then enforce the accuracy of the irradiance-to-power conversion.

- **Nominal physical model**

For benchmarking, a traditional parameter estimation method will also be tested, which extracts the five model parameters from the module manufacture datasheet. Given that the datasheet typically does not explicitly provide the model parameters, the 'pvlib.ivtools.sdm.fit_desoto' [36] function is used for this extraction. This method is named the nominal physical model.

### 2.3.2. Smart persistence model

The persistence model is also a common benchmark model for PV power prediction, which assumes the future power output will closely resemble the historical conditions. The traditional persistence model [37] does not consider the factor of irradiance, which leads to its limited accuracy. Here, we adopt a Smart Persistence model [16] (expressed in (4)), which addresses the forecasted and the historical ground irradiance. Therefore, it can serve as a baseline model with higher accuracy.

$$P(t+h) = P(t)\frac{G(t+h)}{G(t)} \tag{4}$$

where, $G$ refers to the plane-of-array irradiance; $h$ is the prediction horizon (day ahead in this study as 24h).

### 2.3.3. Machine learning models

Machine learning models are widely used in PV power forecasting methods to map the irradiance to output power. Drawing from literature research [9,10], this study employs five common machine learning models: Multilayer Perceptron (MLP) of Artificial neural network (ANN), Random Forest (RF), Support vector regression (SVR), Kernel Ridge (KR), and Linear regression (LR). Input features for the models include the plan-of-array irradiance ($G_{poa}$), module temperature ($T_m$), and hour of day ($HoD$) with the output as the power, as illustrated in Fig. 4.

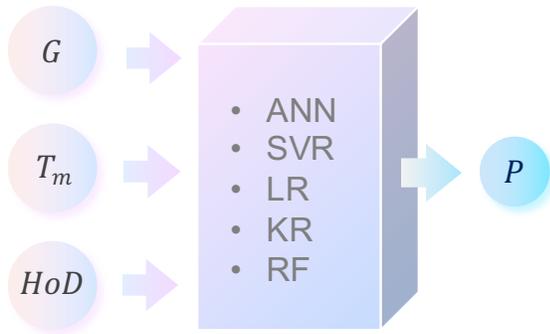

Fig. 4 Flowchart of machine learning models for irradiance-to-power conversion. Input features include irradiance ($G_{poa}$), module temperature ($T_m$), and hour of the day ($HoD$) with the power ($P$) as output. Five models will be tested. Note that these ML models generally require a large amount of historical data for training, which may not be available for newly-installed PV systems.

The performance of machine learning models primarily hinges on the hyperparameters. Based on a comprehensive review of machine learning models for power forecasting [9], a list of the hyperparameters for fine-tuning is outlined in Table S1 of Supplementary Information (SI). The grid search method is used to systematically explore diverse combinations of these hyperparameters. Note that, different from machine learning applications in the literature, we also consider the length of training data as a 'hyperparameter', recognizing that each model has an optimal amount of data for training. We systematically vary the length of training data (measured in 15-minute intervals) from 3 days to 3 months and calculate the power conversion error to determine the most suitable training data length for each model. Further details can be found in Section B of SI.

## 3. Irradiance-to-power conversion performance

This section presents the day-ahead irradiance-to-power conversion results using the various techniques on four field PV systems (system information introduced in Section 2.1.1). To illustrate different case studies, System 1 is selected as the example system to demonstrate the application and the impact of various factors on performance, with the results presented in Section 3.1. The results for all other three PV systems (with different system capacity, location, or PV technology to System 1) are provided in Section 3.2.

### 3.1. Daily irradiance-to-power conversion over a year

The proposed dynamic-model-based technique (PVPro) is applied to System 1 for day-ahead irradiance-to-power, and its performance is compared with the persistence, nominal, and five machine learning models (presented in Section 2.3). An example of predicted power is illustrated in Fig. 5, while the power error spanning an entire year is displayed in Fig. 6.

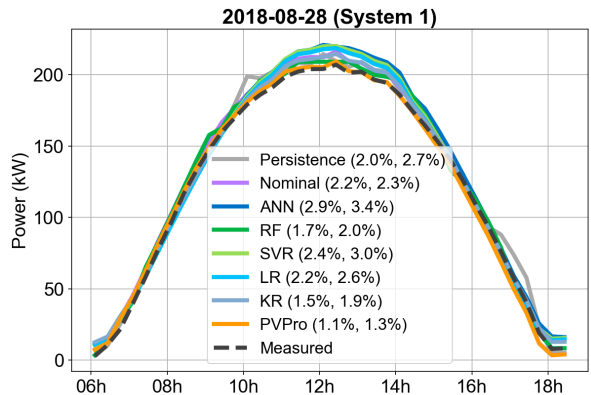

Fig. 5 Example of predicted power on 2018-08-28 of System 1. The legend represents the error metrics as '($nMAE$, $nRMSE$)'. PVPro demonstrates the closest alignment with the measured power throughout the day.

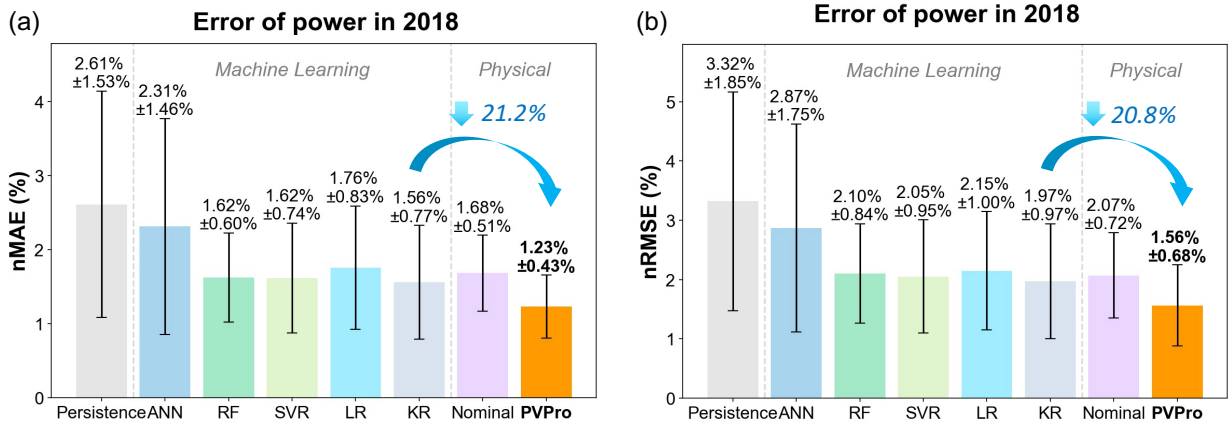

Fig. 6 Errors of the estimated power ((a) $nMAE$, (b) $nRMSE$) using eight techniques on System 1 in 2018. PVPro exhibits lower and more stable errors with a decrease of 21% compared to the best of other techniques (*i.e.*, KR).

From the single-day results in Fig. 5, the power estimated by all the techniques closely follows the overall trend of the measured power, while PVPro exhibits the best alignment with $nMAE$ = 1.1%. As seen from the whole year in Fig. 6, ANN and persistence model display large fluctuation and higher error. PVPro outperforms both the machine learning and physical models by showing a better and more stable performance with a year-averaged $nMAE$ = 1.23% and $nRMSE$ = 1.56%.

### 3.1.1. Effect of seasons

To evaluate the seasonal impact on the irradiance-to-power conversion, we partitioned the year-long results in Fig. 6 into four seasons and calculated the average error, as detailed in Fig. 7. The season windows are defined as follows: Spring (March 1 to May 31), Summer (June 1 to August 31), Fall (September 1 to November 30), and Winter (December 1 to February 28).

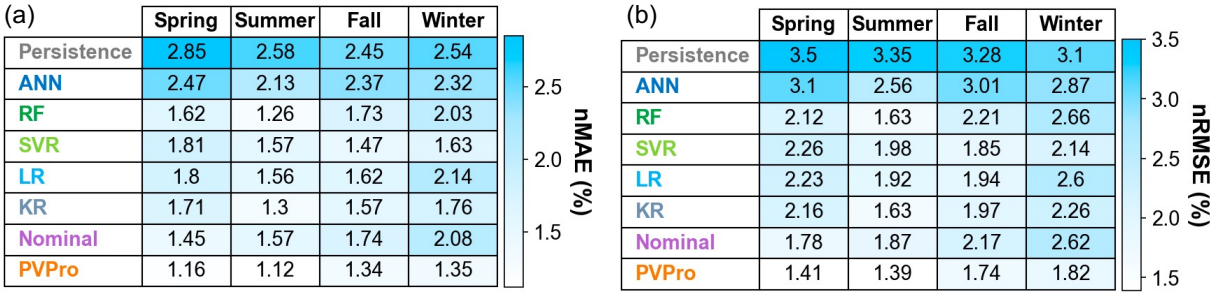

Fig. 7 Errors of the estimated power ((a) $nMAE$, (b) $nRMSE$) under the four seasons in 2018. Overall, the summer is associated with lower power errors, whereas the winter and spring tend to display higher errors. PVPro outperforms other techniques across all seasons.

The performance of each method presents seasonal fluctuations in Fig. 7. Globally, power errors are lower during summer and higher during winter. This may be due to the higher and more stable irradiance in summer, which improves the accuracy of the measurements of environmental conditions and system modeling, and vice versa for winter. Notably, PVPro surpasses other techniques consistently across all seasons.

### 3.1.2. Effect of weather conditions

For data-driven power conversion methods (machine learning, persistence, and PVPro), the performance of the models is directly influenced by the weather conditions in the historical training data. We test 6 typical cases with different weather conditions for the training and test dataset as illustrated in Fig. 8 (a). These cases involve training the models using all clear data, all cloudy data, or a combination of clear and cloudy data, followed by testing with either clear or cloudy data. The clear and cloudy data are selected manually based on the fluctuation of power. The corresponding power errors are depicted in Fig. 8 (b) with the Coefficient of variation ($CV$) value (the ratio of the standard deviation to the mean of the 6 sample values) [38]. These $CV$ values serve to quantify the relative variation of errors across the different weather condition scenarios.

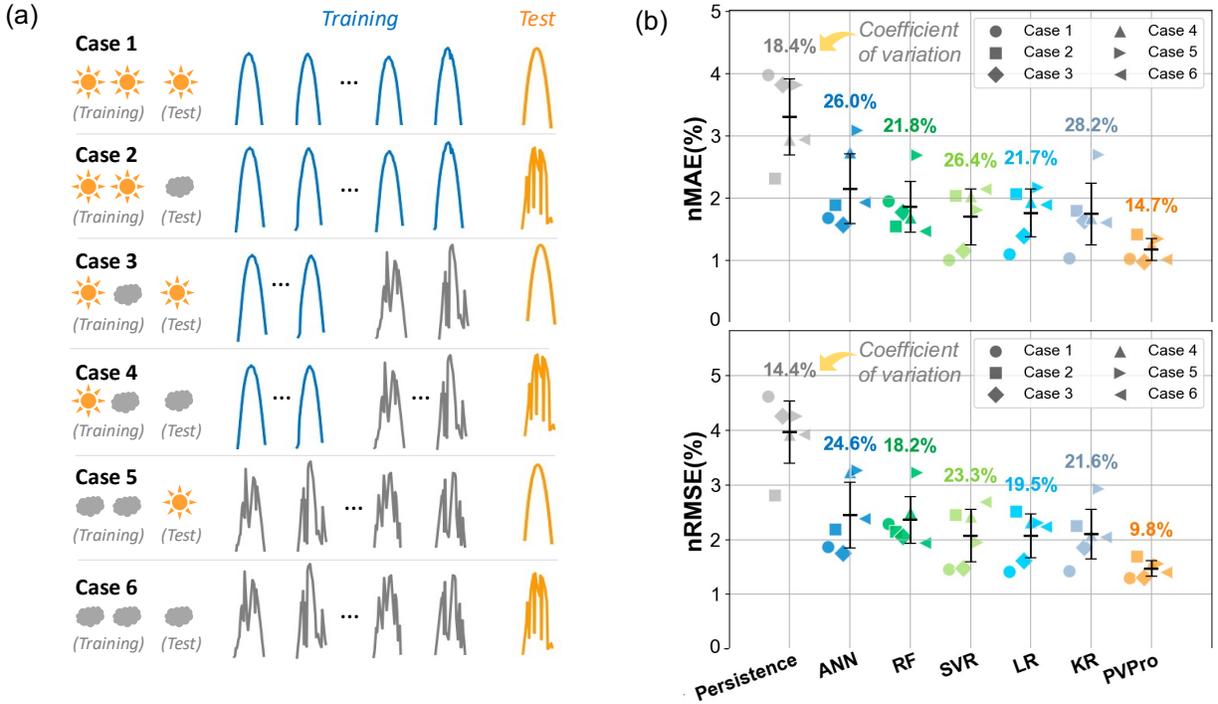

Fig. 8 (a) Six cases with different weather conditions for training (using all clear data, all cloudy data, or a mix of clear and cloudy data) and test (clear or cloudy data) with example field power curve presented. (b) $nRMSE$ and $nMAE$ of converted power under the six cases with the Coefficient of variation ($CV$) value marked on the top of the error point. Compared to other techniques, PVPro shows less fluctuation to the different weather conditions.

From Fig. 8, we can observe a substantial fluctuation of the predicted power error under the 6 cases using ANN, SVR, and KR, with $CV$ >20%. This suggests that these models are more sensitive to variations in the weather conditions present in the training and test data. In contrast, PVPro exhibits less fluctuation across diverse weather conditions, showcasing a higher degree of robustness and stability.

### 3.1.3. Over- and under-estimation analysis

Over- and under-estimation of the output power both risk the instability and reliability issues of the power system. However, the adverse effects are more pronounced with power over-estimation for system operators due to the complexities involved in swiftly deploying backup power units and implementing load reduction measures [37]. Thus, we examine these conditions by quantifying the error using the normalized bias error ($nBE$). The distribution of $nBE$ of predicted power in 2018 is depicted in Fig. 9, where the occurrence frequency (named as density in Fig. 9) for severe over-estimation (when $nBE$ >10% or $nBE$ >20%) is also listed.

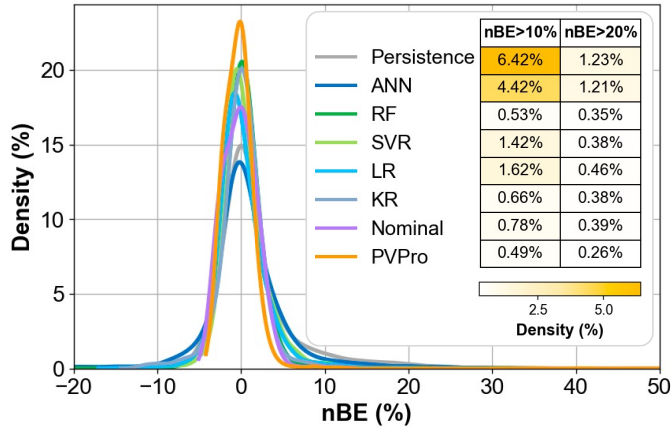

Fig. 9 Distribution of $nBE$ of estimated power in 2018 using different techniques. PV-Pro has a lower frequency (<0.5%) of significant power overestimation ($nBE$ >10% or 20%) compared to other techniques, effectively minimizing the adverse impacts on the power system.

It is shown in Fig. 9 that Persistence and ANN models introduce a higher frequency of severe overestimation (ratio of $nBE$ >10% higher than 4% and ratio of $nBE$ >20% higher than 1%). Comparatively, PVPro achieves the lowest frequency of severe overestimation, underscoring its robust capability to mitigate instances of significant overestimation of the system's power.

### 3.1.4. Interpretability analysis

From the results above, we may notice that, despite all being data-driven methods, PVPro demonstrates superior performance compared to other machine learning models. To further investigate this, we analyze the performance of these models as a function of each input feature. Specifically, for machine learning models, the three inputs are irradiance, module temperature, and the hour of the day (normalized by 24). For each trained model, we systematically vary one feature while keeping all other features constant and plot the estimated power against this variable feature. The results of System 1 in 2018 are depicted in Fig. 10 with a reference line (simulated using datasheet parameters) also plotted (). Note that the reference line may not precisely match the real output of the PV system, but it can provide a theoretical trend of the power against the features. This analysis serves to assess the interpretability of these models.

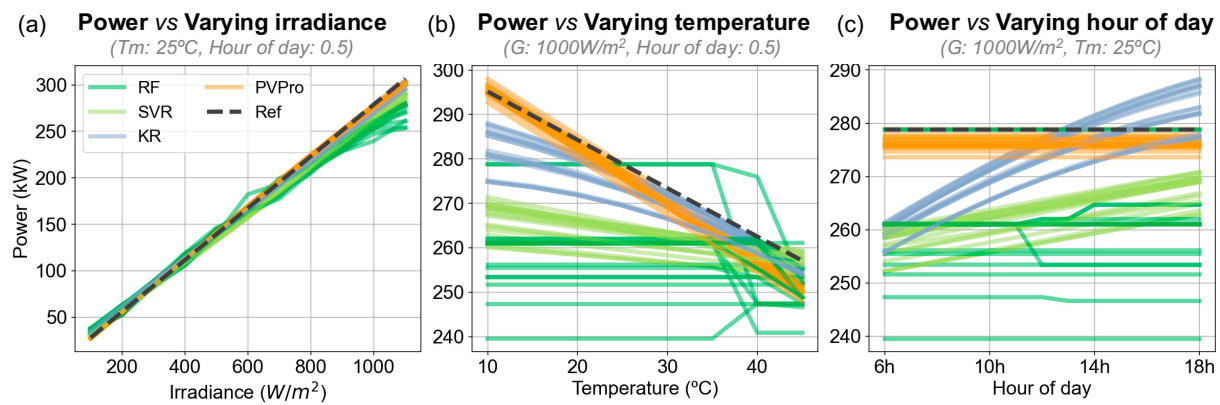

Fig. 10 Estimated power of 30 daily models in June 2018 using input with (a) varying irradiance, (b) varying module temperature, and (c) varying hours of the day (other features are kept constant). ML models follow closely to the reference line in (a), while badly in (b) and (c). The output of PVPro consistently adheres to the reference across all conditions. The ML models lack full interpretability compared to PVPro, particularly in the relationship between power and temperature or hour of the day.

When we vary irradiance and keep others constant (module temperature = 25°C and hour of day = 0.5, *i.e.*, 12h), the output power of all the models globally aligns with the reference trend, *i.e.*, the power increases linearly with the irradiance. However, when the module temperature and hour of the day are varied, significant dispersion occurs among the machine learning models, particularly in ANN and RF. In contrast, PVPro, follows known physical rules of behavior. Thus, it maintains a close alignment with the reference trend and can produce more reliable results. This implies that, even after fine-tuning, machine learning models, still lack complete physical interpretability compared to the physical-model-based approach demonstrated by PVPro.

### 3.2. Application on multiple PV systems

Following the same irradiance-to-power conversion process applied for System 1, Systems 2 to 4 (detailed in Fig. 2) are accordingly analyzed. The $nMAE$ of the predicted error of the four PV systems is shown in Fig. 11, where the '*mean±std*' of the four values of $nMAE$ is marked to show the discrepancy of performance across different systems. Note that for System 4, the module information is unknown. Thus, the nominal physical model cannot be applied to System 4.

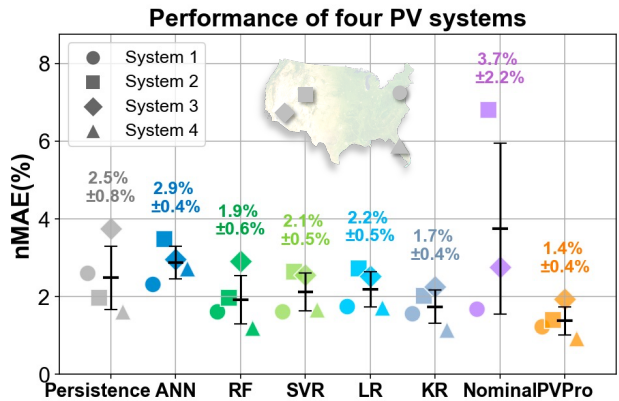

Fig. 11 $nMAE$ of estimated power of four PV systems (presented in Fig. 2). The '*mean±std*' of the four systems $nMAE$ is marked above for each technique. PVPro consistently demonstrates robust performance and lower power errors across different PV systems.

From the results of the four PV systems, it is shown that the model performance varies from site to site. For example, the $nMAE$ of the nominal physical model is below 3% for System 1 and 3 but increases to 6.8% in System 2. This discrepancy is due to the significant degradation of System 2, rendering the nominal parameters unfit for the actual system's condition. The smart persistence model, which relies on the similarity of weather conditions between two consecutive days, also exhibits a distinct performance across different systems.

In contrast, the data-driven models, such as machine learning and PVPro, exhibit an overall less pronounced difference between systems. This is attributed to their ability to adapt and learn from historical production data, allowing them to better capture the system's evolving status. Notably, PVPro exhibits robust performance with lower power errors (average $nMAE$ = 1.4%) across diverse PV systems, reducing the error by 17.6% compared to the best-performing alternative model (*i.e.*, KR).

## 4. Discussion

Various machine learning models were tested in this research, where the selection of models and the fine-tuning of hyperparameters were guided by the successful models reported in the literature. However, across the application of four PV systems with distinct sizes, climate zones, and module technologies, the optimal machine learning model varied from site to site, as presented in Fig. 11. Therefore, we recommend that future researchers interested in machine learning for power prediction not rely solely on a single model recommended in the literature. Instead, we suggest exploring different candidates to identify the most suitable model for the specific PV system. Drawing from this research and literature studies, potential models to evaluate may include kernel ridge (KR), support vector regression (SVR), and random forest (RF).

The persistence model is commonly used as a benchmark for assessing power prediction performance. However, in the literature, many studies employ the *naïve* type of persistence model [39], which simply assumes the future power output will be the same as the past observation. Consequently, this *naïve* model generally leads to relatively poor performance (average $nMAE$ = 15% as shown in Fig. S2 of SI). Using this model as the baseline, the proposed model by researchers often demonstrates more substantial improvements. It is noteworthy that this naïve persistence model could be seamlessly replaced by the smart persistence model (presented in Section 2.3.2) without extra effort. This smart persistence model, considering the impact of irradiance, significantly enhances the accuracy of power predictions (average $nMAE$ = 2.5%), as illustrated in Fig. S2 of SI. Thus, we encourage the use of this updated persistence model as the benchmark for future research.

For the data-driven models, like machine learning and PVPro (fusion of statistical and physical models), the availability of data is crucial. In this research, each machine learning is applied with its optimal length of historical data for training, as discussed in Section 2.3.3. Some models, like ANN, may require up to 60 days of training data for optimal performance. Here, we test an extreme case by providing these models with short-length data (3 days) and present their performance in Fig. S3 in SI. Interestingly, the results indicate that PVPro can still achieve a low prediction error ($nMAE$ = 1.26%) even with this limited training data. This highlights the suitability of PVPro for application in newly-installed PV systems.

Another practical consideration is the computation time. The training time for machine learning and PVPro models, using their optimal length of training data, is within 2 seconds on the Apple MacBook Pro with an M116G chip. This rapid training time fully supports a frequent update (like daily) of model parameters for power prediction.

The Irradiance ($G_{poa}$)-to-power conversion using PVPro is fundamentally a hybridization of statistical (data-driven) and physical methods. The leverage of historical data serves to reconstruct the physical model of the PV system. The power prediction, achieved through equivalent-circuit modeling, ensures that the output power adheres to the physical rules of PV cells and is fully interpretable. As PVPro rebuilds the physical model by fitting the production data, the challenge mainly lies in the quality of the data. Thus, the pre-processing step will be essential, especially in the identification of the operation conditions (inverter on MPP or clipping), removal of outliers, and use of clear-sky data. Improving the pre-processing will be the focus of our next-step work. In essence, the model rebuilt by PVPro, which mirrors the current health status of the PV system, not only enables precise power prediction but also holds promise for operational and maintenance purposes, such as degradation analysis and real-time health monitoring of PV systems.

Finally, the pros and cons of the proposed PVPro and other Irradiance ($G_{poa}$)-to-power conversion models assessed in this paper are summarized as follows:

- **Persistence model**: Simple, highly depends on the temporal correlation of future and past data;

- **Machine learning**: Free of system knowledge, may require a large amount of historical data (up to 60 days), lacks interpretability;

- **Nominal physical model**: Simple, unsuitable for degraded or faulty PV systems;

- **PVPro**: Suitable for degraded PV systems, applicable on new systems with a limited amount of operation data, fully interpretable, robust to seasons and different weather conditions.

## 5. Conclusion

This paper presents a dynamic model-based irradiance-to-power conversion technique (PVPro) for power prediction. PVPro models the day-ahead output power by periodically reconstructing a precise dynamic physical model of the PV system using historical production and weather data. PVPro is compared with the popular irradiance-to-power conversion techniques in the literature. The results reveal that PVPro achieves an outstanding power prediction performance with the average $nMAE$ =1.4% across four field PV systems, on surpassing the best of other techniques with a reduction of error of 17.6%. Additionally, PVPro demonstrates robustness across different seasons and weather conditions present in historical training data. In cases of severe overestimation, PVPro exhibits the lowest frequency, showcasing its effectiveness in mitigating the impact of significant overestimation on the power system. Moreover, PVPro performs well with a limited amount of operational data (3 days), making it suitable for application in newly-installed PV systems. Future work will focus on the improvement of the pre-processing of data and the application of PVPro on more large-scale PV systems. In addition to power prediction, it is also promising to explore its application in real-time health monitoring of PV systems based on the precise system model reconstructed by PVPro. A python-based tool of PVPro for power prediction is available on GitHub: https://github.com/DuraMAT/pvpro.